\documentclass[10pt,conference]{IEEEtran}
\usepackage{cite}

\ifCLASSINFOpdf
  \usepackage[pdftex]{graphicx}
\else
\fi
\usepackage[cmex10]{amsmath}
\begin{document}
%
\title{High-Precision Numerical Simulations of Rotating Black Holes Accelerated by CUDA}

\author{
\IEEEauthorblockN{Rakesh Ginjupalli and Gaurav Khanna}
\IEEEauthorblockA{Physics Department\\
University of Massachusetts Dartmouth\\
North Dartmouth, MA 02747-2300\\
Telephone: (508) 910--6605\\
Email: gkhanna@umassd.edu}}


%


\maketitle

\begin{abstract}
Hardware accelerators (such as Nvidia's CUDA GPUs) have tremendous promise for computational science, because they can deliver large gains in performance at relatively low cost. In this work, we focus on the use of Nvidia's Tesla GPU for high-precision (double, quadruple and octal precision) numerical simulations in the area of black hole physics -- more specifically, solving a partial-differential-equation using finite-differencing. We describe our approach in detail and present the final performance results as compared with a single-core desktop processor and also the Cell BE. We obtain {\em mixed} results -- order-of-magnitude gains in overall performance in some cases and negligible gains in others.  
\end{abstract}


%
\IEEEpeerreviewmaketitle

\section{Introduction}

Computational scientists and engineers have begun making use of hardware accelerators (GPU, Cell BE, FPGA, etc.) because these can provide significant gains in the overall performance of many numerical simulations at a relatively low cost. Compute Unified Device Architecture (CUDA)~\cite{cuda} is NVIDIA's framework for general-purpose computing on its graphics processing units (GPUs); Cell Broadband Engine (Cell BE)~\cite{cell} is a processor that was designed by a collaboration between Sony, Toshiba and IBM and is being used in HDTVs, gaming consoles (Sony's Playstation 3), as well as high-performance computing hardware (IBM's Cell blades~\cite{blades}, LANL RoadRunner~\cite{roadrunner}).

In this work, we make use of these new technologies to accelerate an application from the numerical relativity (NR) community -- a Teukolsky equation solver~\cite{mystuff1,mystuff2,mystuff3,mystuff4}, which is essentially a linear (hyperbolic) partial-difference-equation (PDE) solver code that uses a finite-differencing numerical scheme. A distinguishing aspect of this work is that the numerical simulations presented are such that they require high numerical precision i.e. double (64-bit), quadruple (128-bit) and octal (256-bit) floating-point precision. Therefore, we focus on the implementation of high-precision floating-point arithmetic computation in CUDA, and compare the resulting performance with that from other processor architectures. 

It is worth pointing out that our NR application is of a type that is quite common in various fields of science and engineering, therefore we expect that our work would be of interest to the larger community of computational scientists. These architectures have been recently evaluated for other numerically intensive problems, and their performances have been compared and presented in the relevant literature~\cite{other1,other2,other3,other4,other5}.

This paper is organized as follows: In Section 2, we provide a very brief introduction to CUDA GPUs and also the Cell BE. We emphasize aspects of these hardware accelerators that are relevant to our implementations. In Section 3, we briefly introduce the Teukolsky equation, the relevant background gravitational physics and the numerical method used by the solver code.  It is this code that we accelerate in our work using the Tesla CUDA GPU and also the Cell BE. In Section 4, we describe our parallel code's implementation and then in Section 5 we present the code's overall performance results. Finally, in Section 6, we summarize this work and make some conclusive remarks.

\section{Nvidia CUDA GPU and STI Cell BE} 

All processor manufacturers have moved towards multi-core designs today in the quest for higher performance. At the time of the writing of this article, high-end desktop processors by Intel and AMD have a maximum of six (6) cores. On the other hand, there are other computing technologies that have been in existence for several years that have traditionally had many more compute cores than standard desktop processors. These are sometimes referred to as hardware accelerators and have a {\em many}-core design. Examples of these accelerators include GPUs and the Cell BE. 

The Cell BE~\cite{cell} was developed collaboratively by Sony, IBM and Toshiba primarily for multimedia applications. This processor has a general purpose (PowerPC) CPU, called the PPE (that can run two (2) software threads simultaneously) and eight (8) special-purpose compute engines, called SPEs available for raw numerical computation. Each SPE can perform vector operations, which implies that it can compute on multiple data, in a single instruction (SIMD). All these compute elements are connected to one another through a high-speed interconnect bus (EIB). Note that because of this heterogeneous design, the Cell BE is very different from traditional multi-core processors. A single, 3.2 GHz Cell BE has a peak performance of over 200 GFLOP/s in single-precision floating-point computation and 100 GFLOP/s in double-precision operations. One challenge introduced by this new design, is that the programmer has to explicitly manage the data transfer between the PPE and the SPEs. The PPE and SPEs are equipped with a DMA engine -- a mechanism that enables data transfer to and from main memory and each other. The parallel programming model on Cell BE allows for the use of SPEs for performing different tasks in a workflow ({\em task parallel} model) or performing the same task on different data ({\em data parallel} model). 

In the CUDA context, the GPU (called {\em device}) is accessible to the CPU (called {\em host}) as a co-processor with its own memory. The device executes a function (usually referred to as a {\em kernel}) in a data-parallel model i.e. a number of threads run the same program on different data. The many-core architecture of the GPU makes it possible to apply a kernel to a large quantity of data in one single call. If the hardware has a large number of cores, it can process them all in parallel (for example, Nvidia's Tesla GPU has as many as 240 compute cores clocked at 1.3 GHz). In the area of high-performance computing, this idea of massive parallelism is extremely important. The Tesla GPU can also perform double-precision floating point operations, at a performance comparable to that of the Cell BE mentioned above. GPUs provide significant flexibility in terms of memory management: Six (6) main types of memory exist in the form of registers, local memory, shared memory, global memory, constant memory and texture memory. We will not attempt to go into detail with these different memory arrangements in this document; instead we will simply refer the reader to online resources on this somewhat involved topic~\cite{cuda}.

\section{Numerical Relativity}
Several gravitational wave observatories~\cite{ligo} are currently being built all over the world: LIGO in the United 
States, GEO/Virgo in Europe and TAMA in Japan. These observatories will open a new window onto the Universe by 
enabling scientists to make astronomical observations using a completely new medium -- gravitational waves (GWs), 
as opposed to electromagnetic waves (light). These waves were predicted by Einstein's relativity theory, but have 
not been directly observed because the required experimental sensitivity was simply not advanced enough, until 
very recently.

Numerical relativity~\cite{nr1, nr2, nr3} is an area of computational science that emphasizes the detailed modeling of strong sources 
of GWs -- collisions of compact astrophysical objects, such as neutron stars and black holes. Thus, it plays an 
extremely important role in the area of GW astronomy and gravitational physics, in general. Moreover, the NR 
community has also contributed to the broader computational science community by developing an open-source, modular, 
parallel computing infrastructure called {\it Cactus}~\cite{cactus}.

The specific NR application we have chosen for consideration in this work is one that evolves the perturbations 
of a rotating (Kerr) black hole, i.e. solves the Teukolsky equation in the time-domain~\cite{mystuff1,mystuff2,mystuff3,mystuff4}. 
This equation is essentially a linear wave-equation in Kerr space-time geometry. The next two subsections 
provide more detailed information on this equation and the associated numerical solver code.

\subsection{Teukolsky Equation}
The Teukolsky master equation describes scalar, vector and tensor field perturbations in the space-time of
Kerr black holes~\cite{eqn}. In Boyer-Lindquist coordinates, this equation takes the form
\begin{eqnarray}
\label{teuk0}
&&
-\left[\frac{(r^2 + a^2)^2 }{\Delta}-a^2\sin^2\theta\right]
         \partial_{tt}\Psi
-\frac{4 M a r}{\Delta}
         \partial_{t\phi}\Psi \nonumber \\
&&- 2s\left[r-\frac{M(r^2-a^2)}{\Delta}+ia\cos\theta\right]
         \partial_t\Psi\nonumber\\  
&&
+\,\Delta^{-s}\partial_r\left(\Delta^{s+1}\partial_r\Psi\right)
+\frac{1}{\sin\theta}\partial_\theta
\left(\sin\theta\partial_\theta\Psi\right)+\nonumber\\
&& \left[\frac{1}{\sin^2\theta}-\frac{a^2}{\Delta}\right] 
\partial_{\phi\phi}\Psi +\, 2s \left[\frac{a (r-M)}{\Delta} 
+ \frac{i \cos\theta}{\sin^2\theta}\right] \partial_\phi\Psi  \nonumber\\
&&- \left(s^2 \cot^2\theta - s \right) \Psi = 0  ,
\end{eqnarray}
where $M$ is the mass of the black hole, $a$ its angular momentum per unit mass, $\Delta = r^2 - 2 M r + a^2$ and 
$s$ is the ``spin weight'' of the field. The $s = 0$ versions of these equations describe the radiative degrees 
of freedom of a simple scalar field, and are the equations of interest in this work. As mentioned previously, this equation 
is an example of linear, hyperbolic, homogeneous (3+1)D PDEs which are quite common in several areas of science and 
engineering, and can be solved numerically using a variety of finite-difference schemes.  

\subsection{Teukolsky Code}
Ref.~\cite{klpa} demonstrated stable numerical evolution of Eq.\ (\ref{teuk0}) for using the well-known 
Lax-Wendroff numerical evolution scheme. Our Teukolsky code uses the exact same approach, therefore the contents of 
this section are largely a review of the work presented in the relevant literature~\cite{klpa}. 

Our code uses the tortoise coordinate $r^*$ in the radial direction and azimuthal coordinate $\tilde{\phi}$. 
These coordinates are related to the usual Boyer-Lindquist coordinates by
\begin{eqnarray}
dr^* &=& \frac{r^2+a^2}{\Delta}dr 
\end{eqnarray}
and
\begin{eqnarray}
d\tilde{\phi} &=& d\phi + \frac{a}{\Delta}dr \; . 
\end{eqnarray}  
These coordinates are better suited for performing numerical evolutions in a Kerr space-time background for a number 
of reasons that are detailed in Ref.~\cite{klpa}. Next, we factor out the azimuthal dependence and use the ansatz, 
\begin{eqnarray}
\label{eq:psiphi}
\Psi(t,r^*,\theta,\tilde{\phi}) &=& e^{im\tilde{\phi}} r^3 \Phi(t,r^*,\theta) 
\end{eqnarray}
that allows us to reduce the dimensionality of the PDE to (2+1)D. Defining
\begin{eqnarray}
\Pi &\equiv& \partial_t{\Phi} + b \, \partial_{r^*}\Phi \; , \\
b & \equiv &
\frac { {r}^{2}+{a}^{2}}
      { \Sigma} \; , 
\end{eqnarray}
and
\begin{eqnarray}
\Sigma^2 &\equiv &  (r^2+a^2)^2-a^2\,\Delta\,\sin^2\theta
\; 
\label{pi_eq}
\end{eqnarray} 
allows the Teukolsky equation to be rewritten in first order form as
\begin{eqnarray}
\label{eq:evln}
\partial_t \mbox{\boldmath{$u$}} + \mbox{\boldmath{$M$}} \partial_{r*}\mbox{\boldmath{$u$}} 
+ \mbox{\boldmath{$Lu$}} + \mbox{\boldmath{$Au$}} =  0 ,
\end{eqnarray}
where 
\begin{equation}
\mbox{\boldmath{$u$}}\equiv\{\Phi_R,\Phi_I,\Pi_R,\Pi_I\}
\end{equation}
is the solution vector. The subscripts $R$ and $I$ refer to the real
and imaginary parts respectively (note that the Teukolsky function
$\Psi$ is a complex valued quantity). Explicit forms for the matrices {\boldmath{$M$}},
{\boldmath{$A$}} and {\boldmath{$L$}} can be easily found in the relevant literature~\cite{klpa}.  
Rewriting Eq.\ (\ref{eq:evln}) as 
\begin{equation}
\partial_t \mbox{\boldmath{$u$}} + \mbox{\boldmath{$D$}}
\partial_{r^*} \mbox{\boldmath{$u$}}
=  \mbox{\boldmath{$S$}}\; , 
\label{new_teu2}
\end{equation}
where
\begin{equation}
 \mbox{\boldmath{$D$}} \equiv \left(\begin{matrix}
                    b &   0   &  0  &  0 \cr
                    0  &   b   &  0  &  0 \cr
                    0  &   0   &  -b  &  0 \cr
                    0  &   0   &  0  &  -b \cr
                \end{matrix}\right),
\label{d_matrix}
\end{equation}
\begin{equation}
\mbox{\boldmath{$S$}} = -(\mbox{\boldmath{$M$}} - \mbox{\boldmath{$D$}})
\partial_{r^*}\mbox{\boldmath{$u$}}
- \mbox{\boldmath{$L$}}\mbox{\boldmath{$u$}} 
- \mbox{\boldmath{$A$}}\mbox{\boldmath{$u$}},
\end{equation}
and using the Lax-Wendroff iterative scheme, we obtain stable evolutions.
Each iteration consists of two steps: In the first step, the solution vector 
between grid points is obtained from
\begin{eqnarray}
\label{lw1}
\mbox{\boldmath{$u$}}^{n+1/2}_{i+1/2} &=& 
\frac{1}{2} \left( \mbox{\boldmath{$u$}}^{n}_{i+1}
                  +\mbox{\boldmath{$u$}}^{n}_{i}\right)
- \\
&  &\frac{\delta t}{2}\,\left[\frac{1}{\delta r^*} \mbox{\boldmath{$D$}}^{n}_{i+1/2}
  \left(\mbox{\boldmath{$u$}}^{n}_{i+1}
                  -\mbox{\boldmath{$u$}}^{n}_{i}\right)
- \mbox{\boldmath{$S$}}^{n}_{i+1/2} \right] \; .\nonumber
\end{eqnarray}
This is used to compute the solution vector at the next time step,
\begin{equation}
\mbox{\boldmath{$u$}}^{n+1}_{i} = 
\mbox{\boldmath{$u$}}^{n}_{i}
- \delta t\, \left[\frac{1}{\delta r^*} \mbox{\boldmath{$D$}}^{n+1/2}_{i}
  \left(\mbox{\boldmath{$u$}}^{n+1/2}_{i+1/2}
                  -\mbox{\boldmath{$u$}}^{n+1/2}_{i-1/2}\right)
- \mbox{\boldmath{$S$}}^{n+1/2}_{i} \right] \, .
\label{lw2}
\end{equation}
The angular subscripts are dropped in the above equation for clarity. All angular
derivatives are computed using second-order, centered finite difference expressions. 

Following Ref.~\cite{klpa}, we set $\Phi$ and $\Pi$ to zero on the inner
and outer radial boundaries. Symmetries of the spheroidal harmonics
are used to determine the angular boundary conditions: For even $|m|$
modes, we have $\partial_\theta\Phi =0$ at $\theta = 0,\pi$ while $\Phi =0$ at 
$\theta = 0,\pi$ for modes of odd $|m|$.

\subsection{Kerr Black Hole Tails}

The main science goal for the development of such a Teukolsky equation solver is to study the ``controversial'' Kerr black hole ``tails'' problem. The statement of the problem is simple: place an observer in a circular orbit around a black hole, and have them measure at late times a generic perturbation field, that had compact support at some initial time. It is generally accepted that the observer measures the late-time perturbation field to drop off as an inverse power law of time, specifically as  $t^{-n}$. In the case of a non-rotating black hole, $n=2\ell+3$, where $\ell$ is the multipole moment of the initial perturbation field. Namely, if the initial (compactly supported) perturbation field has the angular dependence of $Y_{\ell}^m$, the angular dependence remains unchanged (due the hole's spherical symmetry) and the decay rate of the field is governed by the $\ell$ value of the initial perturbation. However, in the context of rotating black holes, it is the value of $n$ that has been controversial in the literature, with some conflicting results reported. See for example~\cite{revisited} for a recent and detailed review of the controversy. 

Now, generating accurate numerical simulations in this context involves a number of challenges. Firstly, these simulations need to be rather long -- this is because typically the observed field exhibits an exponentially decaying oscillatory behavior in the initial part of the evolution and only much later this transitions over to a clean power-law decay. Therefore, one needs to wait for the initial oscillations (so called ``quasi-normal ringing'') to dissipate away. Secondly, because each multipole has its own decay rate (which increases with an increase in $\ell$) at late times one ends up with numerical data in which different multipoles have widely different amplitudes (often 30 -- 40 orders of magnitude apart!). For this reason, not only does the numerical solution scheme have to be high-order (to reduce the discretization errors to the required levels) but it also requires high-precision floating-point numerical computation (due to the large range of amplitudes involved). 

Both these requirements make performing scientifically meaningful Kerr tails numerical simulations rather difficult, especially using traditional desktop processors. For this reason, in this work we turn to hardware accelerators such as GPUs and the Cell BE.

\subsection{High-Order and High-Precision}

As mentioned already, we require a higher-order numerical evolution scheme to solve the Teukolsky equation in the context of these Kerr tails simulations. Now, it turns out that it is sufficient that only the angular differentiation (i.e. $\theta$-derivatives of the field) be implemented using a higher-order numerical stencil. The temporal and the radial direction related operations can simply stay 2nd-order and such a mixed approach yields sufficiently good results~\cite{revisited}. For this reason, in this work we choose the finite-difference angular differentiation operator to be 10th-order accurate and leave the rest of the numerical scheme as a standard 2nd-order Lax-Wendroff algorithm. 

In addition, as pointed out before, we also require high-numerical precision -- in particular, double, quadruple and octal precision may be required depending upon the details of Kerr tails simulation being attempted. Now, double-precision (64-bit) floating-point operations are supported on nearly all compute hardware including the Tesla CUDA GPU and the Cell BE. Therefore, no special considerations are necessary for double-precision computations -- we simply use the native double-precision support on each hardware. On the other hand, very few hardware options support quadruple-precision (128-bit) datatype and operations. In fact, amongst the options available to us, only the Cell BE's PPE natively supports {\em long double} datatype with 128-bit accuracy. And to the best of our knowledge, no compute hardware natively supports octal-precision (256-bit) arithmetic. 

Therefore, finding a software solution for our high-precision requirements is necessary. After examining a number of open-source high-precision floating point arithmetic packages, we find that the LBNL QD library~\cite{qd} is one that is well suited for porting over to CUDA and also the Cell BE. In this library, the high-precision datatypes (quadruple and octal precision types) are implementing using a representation based on the appropriate number of double-precision floats and similarly the high-precision floating-point operations are performed ultimately using standard double-precision operations. Figure 1 depicts some sample results from a Kerr tails simulation that makes use of all these enhancements in floating-point precision and also a high-order accurate numerical evolution scheme.   

%
%
\begin{figure}[!t]
\centering
\includegraphics[width=3.5in]{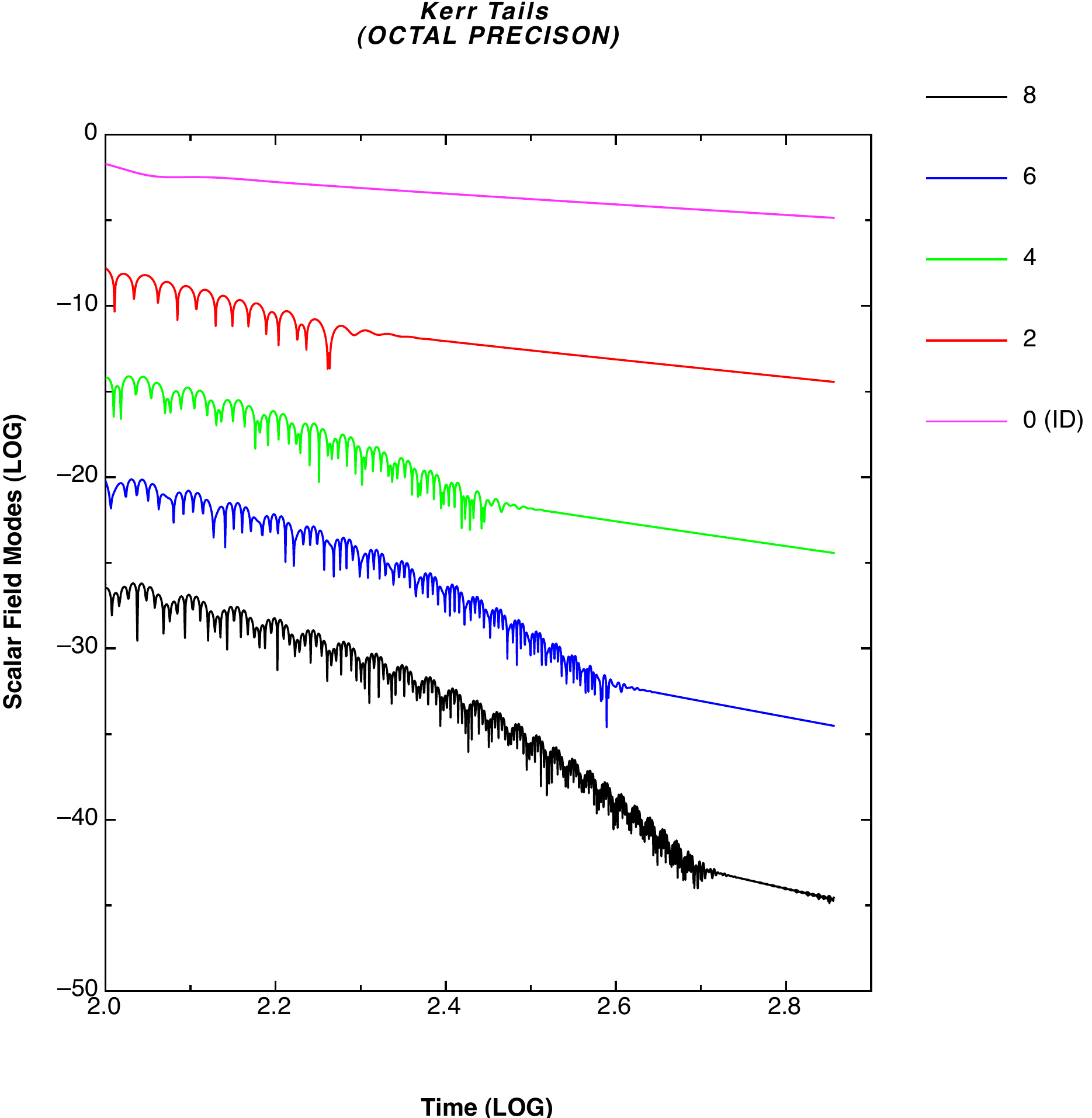}
\caption{Kerr tails for a range of $\ell$ multipoles (0 -- 8) starting with a pure $\ell = 0$ multipole. These tails obey the proposed tail formula $(\ell + \ell' + 3)$~\cite{new}. It is clear that such numerical simulations require octal-precision floating-point arithmetic. } 
\label{tails}
\end{figure}


\section{Implementation Details}

The first task in our work is to isolate the most compute intensive portions of our Teukolsky equation solver code. Upon performing a basic profiling of our code using the GNU profiler {\bf gprof}, we learn that the computing the ``right-hand-sides'' of the Lax-Wendroff steps i.e. the quantities within the square-brackets of Eqs.\ (\ref{lw1}) and (\ref{lw2}), take nearly {\bf 80\%} of the application's overall runtime. We anticipate that this observation is fairly typical for codes of this type. Thus, it is natural to consider accelerating this ``right-hand-side'' computation using data-parallelization on the many-cores of the Tesla GPU or Cell BE. 

A data-parallel model is relatively straightforward to implement in a code like ours. We simply perform a domain-decomposition of our finite-difference numerical grid and allocate the different parts of the grid to different cores. More specifically, on the Tesla GPU, each thread computes the right-hand-side for a single pair of $r^*$ and $\theta$ grid values. In addition, it is necessary to establish the appropriate data communication between the GPU cores and the remaining code that is executing on the CPU -- we use {\bf cudaMemcpy} instructions to transfer data back-and-forth from main memory and we use only global memory on the GPU to simplify communication between the GPU cores. We estimate that this simplification (only making use of global memory) will not impact overall performance significantly because of the relatively intense computation involved in the right-hand-side calculations (especially in the context of the high-precision operations) i.e. the {\em arithmetic intensity} of the computation is sufficiently high. 

Unfortunately, this approach yields negligible performance gain on the Tesla GPU. The reason is that although the right-hand-side computation is accelerated due the use of the many-cores of the GPU, the time it takes to bring that data back/forth to/from main memory so that the remaining computation can resume on the CPU, is large enough that no overall gain in performance is perceived. This is simply due to the poor bandwidth of the system's PCI bus where the GPU is located. It is worth pointing out that this is not an issue for the Cell BE because the general-purpose CPU (PPE) and the many compute-cores (SPEs) reside on the same chip and have an extremely high bandwidth bus (EIB) between them for communication and data transfer. Thus, this approach on the Cell BE nearly yields the maximum allowed performance gain (only limited by {\em Amdahl's Law} to {\bf 5x})~\cite{ijmssc}.

Now, to address this issue on the CUDA GPU, we port {\em all} the Lax-Wendroff related compute routines as separate kernels onto the GPU. In this manner, no communication would be necessary with the rest of the computer system and we would overcome the challenge we face. It is worth noting that some of these routines are not ideal for execution on the GPU (for example, some don't quite have the same level of parallelism that would be essential to obtain high performance from the GPU architecture) but we still port these over for execution on the GPU regardless, simply because our goal is to minimize data transfer back and forth from main memory. This requires a significant amount of additional effort -- but one that pays off well eventually (as seen in the following section). We leave the Cell BE code as outlined before i.e. only accelerate the right-hand-side computation using the SPEs. 

Finally, as mentioned already, we implement high-precision floating-point operations by developing a port of the LBNL QD library for the GPU cores and the Cell's SPEs. In order to do this, we strip out the essentials of the QD representation and the basic functions that we require into separate header files and source-code and make straightforward use of these in our Teukolsky solver code. This approach requires some effort as well, but is not very challenging to implement the details. 

In summary, it is worth pointing out that this high-precision CUDA implementation of our Teukolsky solver code is fairly straightforward, although it does require some effort. It should also be mentioned that we do not attempt to hand-tune the codes to tailor them for each architecture, in order to obtain maximal performance. Instead, we rely on the mature compiler suites to perform all low-level optimizations automatically.

\section{Performance Results}

In this section, we report on the final results from our implementations as outlined in the previous section. We use the following hardware for our performance tests: IBM QS22 blade system, that supports the Cell BE clocked at 3.2 GHz. This system is equipped with 16 GBs of main memory. For the CUDA case, we make use of the Nvidia C1060 Tesla CUDA GPU. This system has an AMD 2.5 GHz Phenom (9850 quad-core) processor as its main CPU and four (4) GBs of memory. All these systems are running Fedora Linux as the primary operating system. Standard open-source GCC compiler suite for code development is available on all these systems.

Table 1 depicts our final performance gains for all the cases considered in this work i.e. double, quadruple and octal precision based parallel implementation of our Teukolsky equation solver code running on CPU, CUDA GPU and Cell BE. The baseline used for this relative comparison is a single-core AMD 2.5 GHz Phenom desktop processor. The Cell BE optimized code performs consistently and expectedly through all the cases considered here. We obtain a performance gain in the range of {\bf 3x -- 5x}, which implies that our parallel implementation has successfully been able to accelerate the right-hand-side computation considerably and now the overall performance is simply limited by the remaining computation that executes serially on the PPE. Naturally, we could improve the performance of our code on the Cell BE even further, by moving more computation onto the SPEs. Interestingly, our CUDA implementation exhibits widely varying performance and we discuss that more thoroughly below. 

%
\begin{table}[!t]
\renewcommand{\arraystretch}{2.5}
\caption{Performance Gain Factor}
\label{table}
\bf
\centering
\begin{tabular}{|c||c||c||c|}
\hline
 & Phenom CPU & Tesla GPU & Cell BE \\
\hline
\hline
Double-Precision & 1x & 20x & 3x \\
\hline
Quad-Precision (QD) & 1x & 1x & 5x \\
\hline
Octal-Precision (QD) & 1x & 4x & 4x \\
\hline
\end{tabular}
\end{table}

\subsection{Double-Precision Performance}

As mentioned before, for the double-precision case we make use the native implementation offered by the hardware itself. This yields a tremendous gain ({\bf 20x}) in overall performance using the Tesla GPU, when compared with a single-core of a typical desktop CPU. Thus, our parallelization approach of moving the entire computation over to the GPU to minimize communication with the CPU pays off very well, even though some of the routines are not ideally suited to run on GPU architectures. 

In this case we certainly obtain an order-of-magnitude gain in overall application performance by making use of a CUDA GPU. 

\subsection{Quadruple-Precision Performance}

For the quadruple-precision case, we make use of our CUDA port of the LBNL QD library in the Teukolsky equation solver code. Interestingly, in this case we obtain negligible gain in overall performance by using the Tesla GPU. However it is worth pointing out if we compare the performance of the QD library based implementation of our code to a {\em long double} based implementation (on the only platform in which that type is available to us -- the Cell BE's PPE), we obtain a {\bf 4x} gain over the latter by simply making use of the QD library.

Thus, when compared with a {\em long double} based implementation, the gains we obtain from all the compute architectures considered in this work are quite significant (nearly {\bf 13x} on the Cell BE).

\subsection{Octal-Precision Performance}

Finally, for the octal-precision case, we also make use of our CUDA port of the LBNL QD library in the Teukolsky equation solver code. In this case, the Tesla CUDA GPU and the Cell BE yield comparable performance gains ({\bf 4x}) over the CPU.  

\section{Conclusions}

In this work, we take an important NR application -- the Teukolsky equation solver code that requires high-precision numerical computation -- and perform low-level parallelization for optimized execution on many-core architectures such as Tesla CUDA GPU and Cell BE. We describe the parallelization approach and its implementation in detail in this article.

The final outcome of our efforts is quite distinct for the different cases considered in this work. On the Cell BE, we obtain a consistent gain in the overall application performance, limited only by the PPE. Thus, there is certainly considerable room for further improvement. On the Tesla GPU, the gains we obtain are negligible in some cases (quadruple-precision) and extremely high in others (double-precision). We believe that the reason behind this is that the smallest ``chunk'' of the computation -- one that would be difficult to parallelize further -- in the context of the high-precision LBNL QD library based computation (quadruple and octal precision) is perhaps still too complex for a single GPU core to compute through efficiently. GPUs are designed to function very well on tasks that can be split into a large number of small, simple and parallel chunks, which is perhaps simply not the case at hand. The LBNL QD library based computation involves significant amount of code branching and irregular memory access patterns -- these are situations that GPUs usually are unable to handle very effectively. Perhaps upcoming CUDA GPU architectures, such as Nvidia's Fermi, will perform significantly better in such situations. 

Our results and observations reinforce the fact that these many-core architectures should not be considered general-purpose computation  accelerators -- the actual computational problem and the relevant parallel approach has to be one that is well suited for the compute hardware's specific design, else one may gain very little or nothing at all. In general, the more massively parallel the task is, the better a design like that of a GPU would perform. For highly serial tasks, a traditional CPU is perhaps still the best option. And finally, for a task that is somewhere in between these two extremes, a design like that of the Cell BE may be optimal.


\section*{Acknowledgment}

We would like to thank Glenn Volkema for his assistance with this work throughout, many helpful discussions and also for providing useful feedback on this manuscript. We are grateful for advice from Jens Breitbart and Anton Obukhov. We would also like to acknowledge research support from the National Science Foundation (NSF grant numbers: PHY-0831631, PHY-0902026), IBM, Sony and Nvidia. 



%

\end{document}